\documentclass[conference]{IEEEtran}
\IEEEoverridecommandlockouts
\usepackage{cite}
\usepackage{amsmath,amssymb,amsfonts}
\usepackage{algorithmic}
\usepackage{graphicx}
\usepackage{textcomp}
\usepackage{xcolor}
\def\BibTeX{{\rm B\kern-.05em{\sc i\kern-.025em b}\kern-.08em
    T\kern-.1667em\lower.7ex\hbox{E}\kern-.125emX}}

\begin{document}

\title{DysLexML: Screening Tool for Dyslexia Using Machine Learning\\
\thanks{\IEEEauthorrefmark{5}Contact author: Maria Papadopouli  (mgp@ics.forth.gr)}
}

\author{\IEEEauthorblockN{
Thomais Asvestopoulou\IEEEauthorrefmark{1}\IEEEauthorrefmark{2},
Victoria Manousaki\IEEEauthorrefmark{1}\IEEEauthorrefmark{2},
Antonis Psistakis\IEEEauthorrefmark{1}\IEEEauthorrefmark{2}, \\
Ioannis Smyrnakis\IEEEauthorrefmark{2}\IEEEauthorrefmark{3}, 
Vassilios Andreadakis \IEEEauthorrefmark{3},
Ioannis M. Aslanides\IEEEauthorrefmark{4} and
Maria Papadopouli\IEEEauthorrefmark{1}\IEEEauthorrefmark{2}\IEEEauthorrefmark{5}\\
\IEEEauthorblockA{\IEEEauthorrefmark{1}Department of Computer Science, University of Crete, Heraklion, Greece\\
\IEEEauthorrefmark{2}Institute of Computer Science, Foundation for Research and Technology-Hellas, Heraklion, Greece\\
\IEEEauthorrefmark{3}Optotech Ltd., Heraklion, Crete, Greece \\
\IEEEauthorrefmark{4}Emmetropia Eye Institute, Heraklion, Greece \\
}}}

\maketitle

\begin{abstract}
Eye movements during text reading can provide insights about reading disorders. Via eye-trackers, we can measure when, where and how eyes move with relation to the words they read. Machine Learning (ML) algorithms can decode this information and provide differential analysis. 
This work developed DysLexML, a screening tool for developmental dyslexia that applies various ML algorithms to analyze fixation points recorded via eye-tracking during silent reading of children. It comparatively evaluated its performance using measurements collected in a systematic field study with 69 native Greek speakers, children, 32 of which were diagnosed as dyslexic by the official governmental agency for diagnosing learning and reading difficulties in Greece.
We examined a large set of features based on statistical properties of fixations and saccadic movements and identified the ones with prominent predictive power, performing dimensionality reduction. Specifically, it achieves its best performance using linear SVM model, with an accuracy of 97\%, over a small feature set, namely saccade length, number of short forward movements, and number of multiply fixated words. 
Furthermore, we analyzed the impact of noise on the fixation positions and showed that DysLexML is accurate and robust in the presence of noise. These encouraging results set the basis for developing screening tools in less controlled, larger-scale environments, with inexpensive eye-trackers, potentially reaching a larger population for early intervention.

\end{abstract}


\section{Introduction}\label{introduction}
Dyslexics manifest significant and persistent reading difficulties ~\cite{b6}, which often involve difficulty in reading due to word decoding (relating sounds with written phrases, i.e., graphemes to phonemes) \cite{b11}. Early intervention can be effective in alleviating the symptoms of the disability. However, screening large populations of children is rather time-consuming and expensive \cite{b17}. For example, a differential diagnosis of dyslexia can take up to 14 months \cite{b1}.
It has been known that the eye movements during text reading can be particularly revealing \cite{b12,b13,b14,b16,b19}. For example, dyslexics exhibit more aberrant eye movements than normal readers at the same age level \cite{b6}, although it is unlikely that the primary cause of dyslexia is erratic eye movements.
{\em Fixations}, i.e., maintaining the visual gaze on a single location, and {\em saccadic movements}, i.e., quick simultaneous movements of the eyes between fixations, are important characteristics for screening dyslexia.
Readers with developmental dyslexia generate different eye movements than typical readers during text reading: longer and more frequent fixations, shorter saccade lengths, more backward refixations than typical readers \cite{b7,b13,b14,b16}.
Furthermore, readers with dyslexia have difficulty in reading long words, lower skipping rate of short words, and high gaze duration (total fixation duration at first visit) on many words. Nonetheless, it is still an open question whether it is possible to build a screening tool that can reliably identify readers who may be of high risk for dyslexia by analyzing these distinctive oculomotor patterns collected during reading and can be robust under noise. \par
This work develops DysLexML, a screening tool for dyslexia, that employs various ML classifiers, such as SVM, Na\"{\i}ve Bayes, and comparatively evaluates their performance using data collected in the field study of RADAR\cite{b1}. To examine its robustness, we assessed its accuracy under various fixation position noise levels, introduced by the eye-tracking technology or the small screen size (e.g., the small size of the text when a mobile device is used).
For that, Gaussian noise of increasing standard deviation is added to the fixation positions of the dataset.
This work demonstrates that DysLexML can achieve high accuracy (of 97\%) and is robust in the presence of noise. 
It performs dimensionality reduction, achieving the aforementioned performance using {\em only a small number of features}, namely the mean and median saccade length, number of short forward movements, and number of multiply fixated words.

The innovative contributions of this work are the analysis of the robustness in the presence of noise, the high accuracy using only a small set of features, and the comparative evaluation with other screening tools/algorithms.
The paper overviews briefly the field study in Section \ref{background}. Section \ref{system} presents the DysLexML and Section \ref{perf} evaluates its performance. Section \ref{conclusion} summarizes our key findings and future work plans.
\section{Background and Field Study}\label{background}
The field study of RADAR was performed in Greece and included 69 children, 32 of which were diagnosed as dyslexic by the official governmental agency for diagnosing learning and reading difficulties in Greece. Participants’ age span is between 8.5 and 12.5 years old. 
The children were instructed to read two passages, at their own pace. It was also emphasized that the purpose was to understand the texts in order to answer five comprehension questions at the end. 
Both texts were written by a special education teacher in Greek. The first passage ({\em baseline} text) consists of 181 words, many of which multi-syllable. A second passage, simpler than the first one, targeting to younger participants, was also given to the subjects. It included 143 words, mostly of one or two syllables. The experimental procedure consisted of recording the eye movements of the participants, while they were silently reading the texts in front of a computer monitor.

A custom-made eye-tracker, developed by Medotics AG was employed. It consists of two steady cameras that can record images up to 60Hz with a resolution of 1600$\times$1200 pixels. Cameras are positioned between the screen and participant with a viewing field from down towards the participant's face. While the participant performs a reading task, the cameras record the participant's face. The images extracted are then used to detect pupil and corneal reflection coordinates.
Based on the collected raw gazing measurements, the fixations were identified according to a dispersion algorithm \cite{b21}.
More information about the field study, e.g., the inclusion and exclusion criteria, texts, and data collection, can be found in \cite{b1}.
A dataset that includes for each fixation, its x- and y-axis coordinates, its starting and termination time, as well as the Region of Interest (ROI) (i.e word) the subject is looking at, is provided as input to DysLexML for analysis.

\begin{figure}[h]
\centerline{\includegraphics[width=\columnwidth]{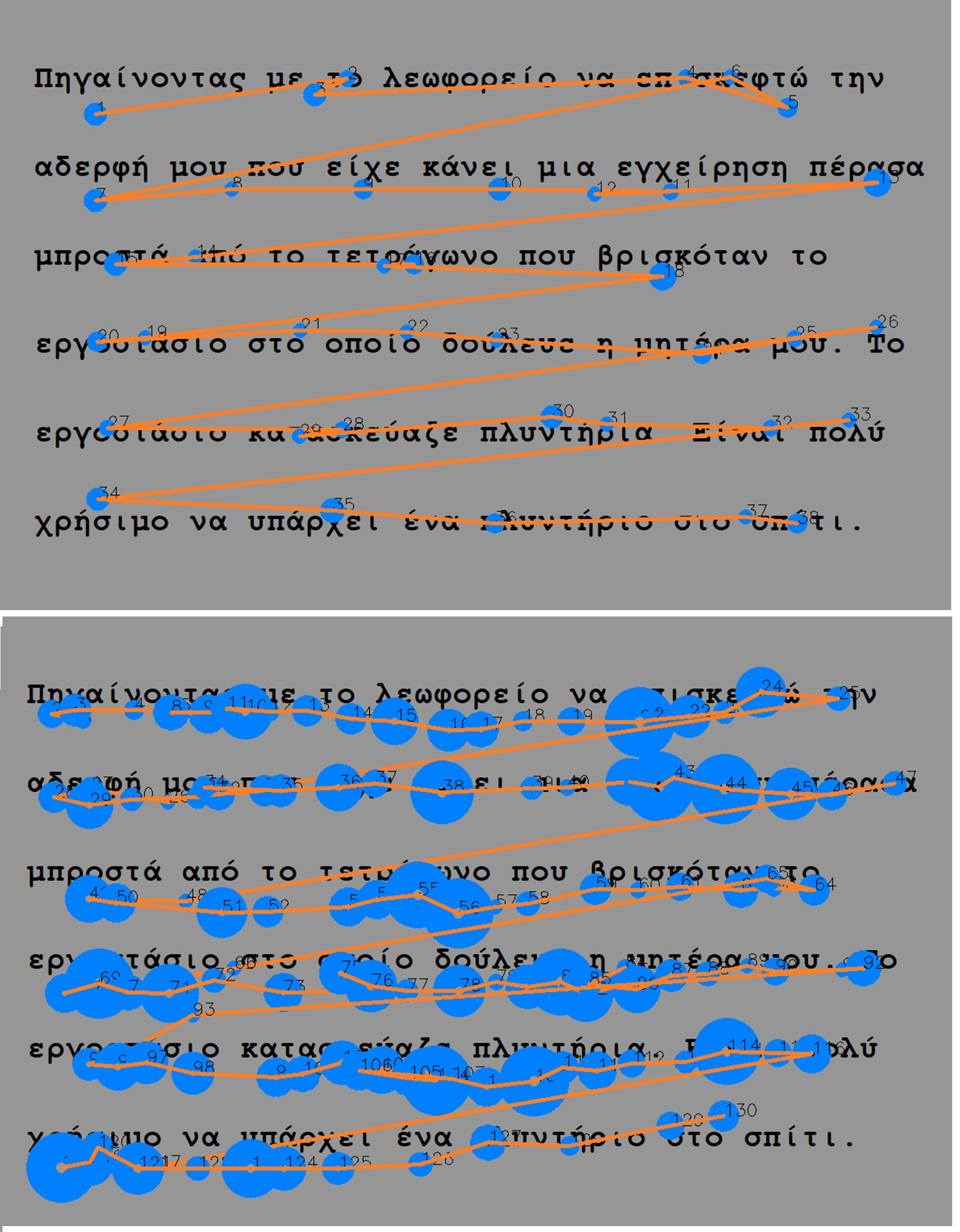}}
\caption{Reading "path" from a typical reader (top) and from a reader with dyslexia (bottom). The blue circles are the fixations and the orange lines the saccadic movements. The larger the circle, the longer the fixation (Figure appeared in \cite{b1}).}
\label{fig0}
\end{figure}

\section{DysLexML system}\label{system}
The main modules of the DysLexML algorithm include the feature extraction, the feature selection for identifying the dominant features, and its classifiers that employ these dominant features.
DysLexML extracts {\em general (non-word-specific)} features and {\em word-specific} ones that take into account the word the subject is looking at. Examples of non-word specific features are the number of fixations on the screen, mean and median duration of fixations and related to saccades, the mean and median length of saccades, i.e., the Euclidean distance between consecutive fixations, and characterization of the types of eye movements.
DysLexML creates a feature vector of 35 features in total.

People with reading difficulties tend to perform back and forth movements (saccades) on the text line as they proceed as a result of difficulty to focus or understand \cite{b13}. Thus the identification of such movements and definition of features based on them can provide valuable information about the dyslexic population. 
Typical readers tend to perform medium to large movements (saccades) in terms of length, while readers with reading difficulties "generate" many choppy movements\cite{b7}.
A movement is labeled as {\em short} if the Euclidean distance between its consecutive fixations is less than 100 pixels (about 5 letters in the text). Most of the movements occur within words. Given that the line of text was about 900 pixels, the threshold for {\em medium to long movements} was set to be 400 pixels (about half a line). With this threshold a {\em medium backward movement} includes re-reads of small groups of words but not of entire phrases. That is, short movements are of less than 100 pixels, long ones are of more than 400 pixels, and medium movements in the range between 100 and 400 pixels. Change-of-line movements have been excluded from both forward and backward movement sets. We also derive information about the number of visits of each word, namely the number of words that were not visited at all (skipped) and number of words that where visited more than once during the text reading.
To identify the features with the most predictive power, we employed the least absolute shrinkage and selection operator (LASSO) \cite{b8}, a particular case of penalized least squares regression with L1-penalty function. LASSO finds the minimum of the residual sum of squares, subject to the sum of the absolute value of the coefficients being less than a constant.
The LASSO estimate can be defined by:
\begin{equation}
\centerline{\includegraphics[width=\columnwidth]{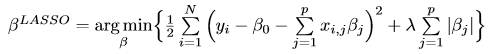}}
\end{equation}

\begin{figure}[h]
\centerline{\includegraphics[width=\columnwidth]{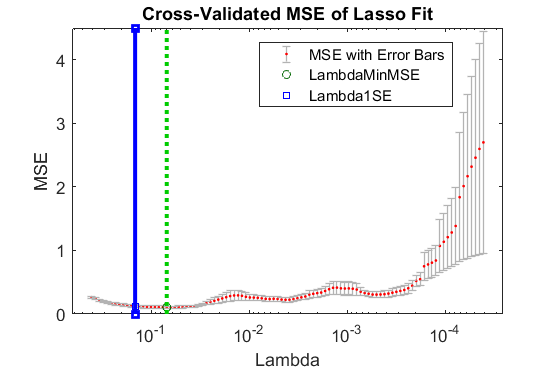}}
\caption{Cross-Validated MSE of Lasso Fit for the baseline text.}
\label{fig1}
\end{figure}

In practice, as $\lambda$ gets higher, less features are taken into account. Specifically, the parameter $\lambda$ in LASSO regression is estimated using 5-fold cross validation. Two values of $\lambda$ were examined, namely the $\lambda_{minMSE}$ that corresponds to the minimum mean cross-validation error MSE (vertical dotted line in Fig.~\ref{fig1}) and $\lambda_{1SE}$ which is one standard error of the mean higher than $\lambda_{minMSE}$ (vertical solid line). 
The purpose of the addition of the 1SE is to reduce the number of regression coefficients, while the mean square error remains close enough (1SE) to $\lambda_{minMSE}$.
Both variations were considered in our analysis.

DysLexML builds classifiers based on SVM, Na\"{\i}ve Bayes, and K-means. The SVM with a linear kernel performs better than the ones with Gaussian or Polynomial, so only the performance of the linear kernel is reported here. The K-Means-based classifier was built as follows: the subjects of the training set are clustered using k-means, and a label is assigned to each cluster based on the most frequent label within that cluster. The distance of the test subject from the centroid of each clusters was estimated. The classifier reports the label of the cluster whose centroid has the shortest distance from the test data.
\section{Performance analysis}\label{perf}
DysLexML consists of two phases: It first employs the LASSO Regression five-fold cross-validation to identify the dominant features. Based on the dominant features, it applies various classification algorithms. For evaluation, RADAR uses the Leave One Out Cross validation (LOOCV), an appropriate choice given the relatively small size of the dataset. 
To comparatively evaluate DysLexML with RADAR, we used LOOCV and the same subject populations. Given that there were subjects with missing values in the word specific features, we filled in the missing values with the median of the corresponding feature values of the training set.
\begin{table}[htbp]
\caption{Classification performance, including all subjects, treating the missing values. The first column corresponds to the baseline text, while the second column to the easier text.}
	\begin{center}
	\begin{tabular}{|c|c|c|}
		\hline
		Classifier & \multicolumn{2}{|c|}{LOOCV accuracy} \\
		\hline
		{K-means (k=2) , LASSO ($\lambda_{minMSE}$)} & 86.95 & 89.39 \\
	    {K-means (k=3) , LASSO ($\lambda_{minMSE}$)} & 91.30 & 84.84 \\
		{K-means (k=4) , LASSO ($\lambda_{minMSE}$)} & 81.15 & 84.84 \\
		{K-means (k=2) , LASSO ($\lambda_{1SE}$)} & 89.85 & 78.78 \\
		{K-means (k=3) , LASSO ($\lambda_{1SE}$)} & 86.95 & 84.84 \\
		{K-means (k=4) , LASSO ($\lambda_{1SE}$)} & 89.85 & 83.33 \\
		{Linear SVM, LASSO ($\lambda_{minMSE}$)} & 94.20 & 80.30 \\
		{Linear SVM, LASSO ($\lambda_{1SE}$)} & 97.10 & 87.87 \\
		{Linear SVM, without feature selection} & 85.50 & 81.81 \\
		{Na\"{\i}ve Bayes, LASSO ($\lambda_{minMSE}$)} & 91.30 & 86.36 \\
		{Na\"{\i}ve Bayes, LASSO ($\lambda_{1SE}$)} & 92.75 & 84.84 \\
		Trivial Accuracy & 53.62 & 53.03 \\
		\hline
	\end{tabular}
	\label{table1}
\end{center}
\end{table}

The exclusion of the word-specific features from the feature vector results to a lower average accuracy for the baseline (difficult) text. The performance remains the same in the case of the easier text, indicating that the word-specific features are not useful when the text is not challenging for the reader.

DysLexML, with SVM and LASSO ($\lambda_{1SE}$), outperforms RADAR: 97.10 \% vs. 94.2 \% for the baseline text (Table.~\ref{table1}). For the easy text, RADAR reports 87.9 \% correct classification, while DysLexML, with K-means with k equal to 2, exhibits an accuracy of 89.39 \%.

\begin{figure}[h]
\centerline{\includegraphics[width=\columnwidth]{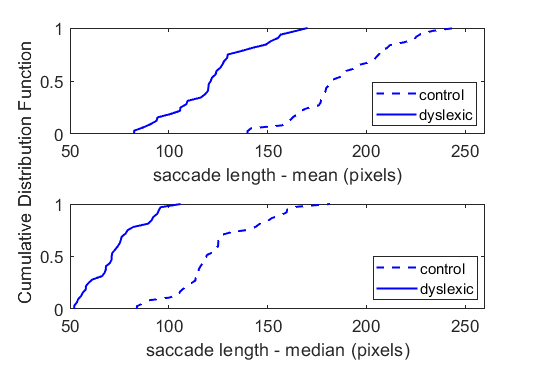}}
\caption{ECDF of saccade length features.}
\label{fig2}
\end{figure}

The dominant features (as selected by the LASSO) for both texts are the mean saccade length and number of short forward movements. In the case of the baseline (difficult) text, the additional dominant features are the median saccade length, and the number of multiply fixated words. 
Prior research has also reported the important role of these dominant features identified by LASSO (as discussed in Section \ref{introduction}).
The distributions of the mean and the median saccade length for both populations are significantly different (as shown in Fig.~\ref{fig2}), which explains their presence as separate dominant features.

\begin{figure}[htbp]
\centerline{\includegraphics[width=\columnwidth]{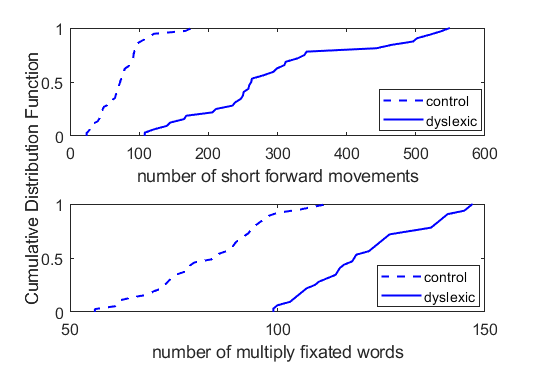}}
\caption{ECDF of number of short forward movements (top) and number of multiply fixated words, i.e. the words that have been fixated more than once during the reading session (bottom).}
\label{fig3}
\end{figure}

Note that there is diversity in the dyslexic population: not all cases are equally severe. Moreover, remember the subjects were instructed to not rush their reading and understand the text in order to answer some comprehension questions at the end. This may have prolonged the reading sessions even for typical readers.
The number of short forward movements was expected to play a prominent role. Dyslexics have been reported to perform more and shorter saccades during reading, in their attempt to decode the text \cite{b14}. The number of short forward movements of the dyslexic population has large variance, as shown in the upper part of Fig.~\ref{fig3}. 50 \% of the dyslexic subjects have more than twice total short progressive movements than the control population.

Dyslexics tend to revisit words more, especially those that are long or difficult to read \cite{b5}. 90 \% of the typical readers have less than 100 words fixated more than once, while this is the starting value for the dyslexic subjects (Fig.~\ref{fig3} (bottom)). This illustrates the value of the word specific analysis of the eye-tracking study.

\begin{figure}[htbp]
\centerline{\includegraphics[width=\columnwidth]{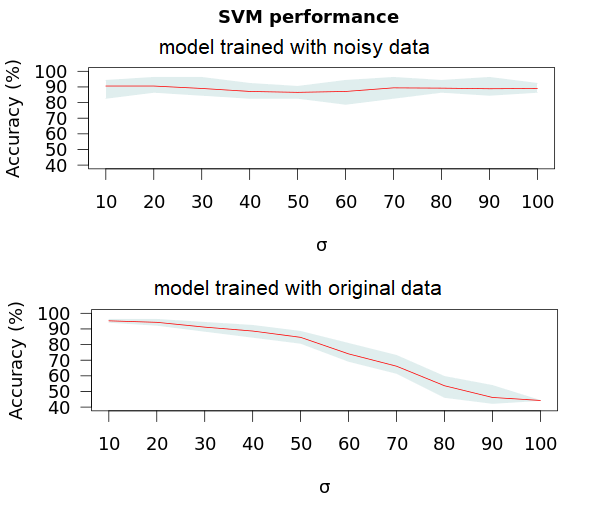}}
\caption{Performance of SVM model under noise. Training model with noisy data (top) and with original data (bottom). The testing was performed on noisy data (10 for each $\sigma$ value). The solid red line indicates the mean accuracy over the 10 noisy synthetic datasets, while the gray area represents the range between the lowest and the maximum accuracy achieved at each noise level.}
\label{fig4}
\end{figure}

To examine the robustness of DysLexML, we evaluated its performance in the presence of noise in the form of small displacements of the fixation points. For this analysis, we considered only the children that had reliable data (according to \cite{b1}) and no missing values. The noise follows a Gaussian distribution with mean value equal to zero and standard deviation varying from 10 to 100 pixels (with a step size of 10). 
In the case of small displacement, only the saccadic movement features changed. 
However, large displacements result to significant changes in the word-specific features. For each subject, the noise was added and the new feature vectors were generated. Note that the “shifted” eye-movements result to different feature vectors. The DysLexML was then evaluated for this new dataset.
Specifically, for each $\sigma$, we generated 10 synthetic datasets. We run the linear SVM LOOCV with LASSO ($\lambda_{1SE}$) feature selection on 100 datasets. DysLexML is robust under noise (Fig.~\ref{fig4} (top)).
We then trained a linear SVM model using the dominant features that were reported by LASSO using the {\em original} dataset. For testing, we employed the 10 synthetic datasets with noise for each given $\sigma$. Fig.~\ref{fig4} (bottom) presents the acquired results. The model exhibits a robust performance for relatively small noise levels (up to $\sigma$ of 30 pixels), which corresponds to about 1 character on the x-axis and 1/3 of the line on the y-axis. However, as the noise level increases, the accuracy drops significantly.
DysLexML through the SVM, that behaves well under generalization, addresses the noise in the fixation coordinates in a robust manner.
 
\section{Related Work}\label{related_work}
Although dyslexia has been extensively studied the last three decades with specialized eye-trackers, there is only a limited number of eye-tracking-based screening systems, partially due to the high cost of eye-trackers up to recently and the debate of the primary cause of dyslexia \cite{b16}. The lack of extensive datasets limits significantly the performance of deep-learning architectures. On the other hand, SVM is powerful in case of relatively small datasets. Recent studies have applied ML, and more specifically SVM for classification of dyslexia on data collected from eye-trackers \cite{b2,b3}. 
For example, Rello and Ballesteros \cite{b2} performed a field study that included 97 Spanish language native speakers, aged 11-54 reading 12 different texts. They used a binary polynomial SVM classifier and achieved classification accuracy of 80.18\%. Their feature vector included the age of the participant, the text number, details about text stylistics, number of visits of a ROI, mean time spent on a ROI, total reading time, mean of fixation duration, number of fixations and sum of all fixation durations. They reported that the reading time, the mean of fixation duration, and the age of the participant have predictive power.
Benfatto {\em et al.} \cite{b3} also employed linear SVM with sequential optimal optimization for screening dyslexia. Their field study in Sweden included 185 children, 97 of them with high risk of dyslexia, speaking Swedish as a first language. All the subjects were reading from paper a short text adapted to their age while their eye movements were recorded. The subjects were equipped with head-mounted goggles with arrays of infrared transmitters and detectors, arranged around each eye. A chin and forehead rest were deployed to minimize head movements and stabilize the viewing distance. Their feature set, produced using a dynamic dispersion threshold algorithm, consisted of 168 features.
They also distinguished saccades to progressive and regressive ones. A  recursive feature elimination algorithm identified the dominant features. They achieved accuracy of 95.6\%$\pm$4.5\% using 48 features of the original feature space.
Al-Edaily {\em et al.} \cite{b4} developed “Dyslexia Explorer” in the Arabic language and performed a study with 14 subjects, 7 of whom with diagnosed dyslexia. Their system is designed to help specialists analyze visual patterns of reading and provide insights into understanding differences between readers with and without dyslexia. Their measurements included fixation duration in each/ all ROI, mean fixation duration in each/ all ROI, total fixation count for each/ all ROI and backward saccades. 
Unlike the above ML-based approaches, Smyrnakis {\em et al.} \cite{b1} developed statistical Bayesian classifiers, using various thresolds and taking into consideration binary correlations. They focused on small age span, critical for dyslexia diagnosis. The size and font of the two texts used was standardized so as to achieve maximal classification accuracy, unlike in \cite{b2}. The parameters used for classification involved not only direct eye-tracking parameters, but also relations between eye-tracking parameters and word properties in the texts read. These parameters extend the parameter set used in \cite{b3}. Including this set of parameters, it is possible to evaluate word anticipation, which is often problematic in dyslexics \cite{b20}. 
Our work employs the same dataset as in \cite{b1}. However, DysLexML applies and evaluates various ML classifiers. As mentioned, the classifier with the best accuracy on noise-free data is the linear SVM classifier on features selected by the LASSO regression at $\lambda_{1SE}$. Furthermore, it exhibits a robust performance under fixation position noise (added artificially). Its robustness and ability to perform dimensionality reduction are the two innovative aspects of this work.
\section{Conclusion}\label{conclusion}
Feature selection, here via LASSO with $\lambda$ of 1 standard error, enabled dimensionality reduction, without compromising the accuracy. 

The mean and median saccade length, the number of short forward movements, and the number of multiply fixated words are the four features with the most prominent predictive power for the baseline text, while for the easier text only the mean saccade length and the number of short forward movements were selected. 
The text difficulty does play an important role in the diagnosis: Easy, less challenging, text, reduces the power of the word-specific features, as they do not appear in the dominant feature set. The text choice has to be relevant to the subjects’ age and so far acquired reading skills. The selected features are easily interpreted and capture the prior knowledge about eye movements of dyslexic children. To the best of our knowledge, DysLexML uses the smallest feature set, compared to the other related studies. 
We envision the development of a system that can operate in a less controlled, larger-scale environment (e.g., potentially in kindergartens or homes) with commercial eye-trackers, reaching a larger population. As a first step towards this objective, here we added synthetic noise at the fixation positions and assessed its impact on the accuracy. For noise levels smaller than $\sigma$ equal to 40 pixels the performance of the system remains robust.
Encouraged by the robustness under noise, the team has performed a follow-up larger-scale field study using inexpensive non-specialized eye-trackers in a more diverse setting (in different countries and under silent and out-loud reading). We aim to identify the different classes of reading difficulties. This work sets the basis for developing a screening tool that can reach a larger more diverse population, in less controlled environments, for early intervention and potentially larger social impact.

\bibliographystyle{unsrt}

\end{document}